\documentclass[3p,times,twocolumn]{elsarticle}

\usepackage{ecrc}

\volume{00}

\firstpage{1}

\journalname{Nuclear Physics B Proceedings Supplement}

\runauth{}

\jid{nuphbp}

\jnltitlelogo{Nuclear Physics B Proceedings Supplement}

\usepackage{amssymb}

\usepackage[figuresright]{rotating}

\begin{document}

\begin{frontmatter}



\dochead{}

\title{Induced magnetic moment in the magnetic catalysis of chiral symmetry breaking}


\author{Efrain J. Ferrer and Vivian de la Incera}

\address{Department of Physics, The University of Texas at El Paso
500 W. University Ave., EL Paso, TX 79968}

\begin{abstract}
The chiral symmetry breaking in a Nambu-Jona-Lasinio effective model of quarks in the presence of a magnetic field is investigated.  We show that new interaction tensor channels open up via Fierz identities due to the explicit breaking of the rotational symmetry by the magnetic field. We demonstrate that the magnetic catalysis of chiral symmetry breaking leads to the generation of two independent condensates, the conventional chiral condensate and a spin-one condensate. While the chiral condensate generates, as usual, a dynamical fermion mass, the new condensate enters as a dynamical anomalous magnetic moment in the dispersion of the quasiparticles. Since the pair, formed by a quark and an antiquark with opposite spins, possesses a resultant magnetic moment, an external magnetic field can align it giving rise to a net magnetic moment for the ground state. The two condensates contribute to the effective mass of the LLL quasiparticles in such a way that the critical temperature for chiral symmetry restoration becomes enhanced.
\end{abstract}


\end{frontmatter}


\section{Introduction}
\label{Sec-1}

The phases of matter under strong magnetic fields constitute an active topic of interest in light of the experimental production of large magnetic fields in heavy-ion collisions, and also because of the existence of strongly magnetized astrophysical compact objects. On the other hand, from a theoretical point of view there exist contradictory results about the influence of a magnetic field on the chiral and deconfinement transitions of QCD \cite{QCD-latticeB}.

Of particular interest for the present paper are some recently obtained results \cite{NJL-B} on the influence of a magnetic field on the condensate structures characterizing the QCD chiral transition. A magnetic field is known to produce the catalysis of chiral symmetry breaking (MC$\chi$SB) \cite{MC} in any system of fermions with arbitrarily weak attractive interaction. This effect has been actively investigated for the last two decades \cite{severalaspects}. In the original studies of the MC$\chi$SB, the catalyzed chiral condensate was assumed to generate only a dynamical mass for the fermion. Recently, however, it has been shown that in QED \cite{ferrerincera}  the
MC$\chi$SB leads to a dynamical fermion mass and inevitably also to a dynamical anomalous magnetic moment (AMM). This is connected to the fact that the AMM does not break any symmetry that has not already been broken by the other condensate. The dynamical AMM in massless QED leads, in turn, to a non-perturbative Lande g-factor and Bohr magneton proportional to the inverse of the dynamical mass. The induction of the AMM also gives rise to a non-perturbative Zeeman effect \cite{ferrerincera}. An important aspect of the MC$\chi$SB is its universal character for theories of charged massless fermions in a magnetic field. Therefore, it is naturally to expect that the dynamical generation of the AMM shall permeate all the models of interacting massless fermions in a magnetic field. 

As follows, we consider the dynamical generation of a net magnetic moment in the ground state of a one-flavor Nambu-Jona-Lasinio (NJL) model in a magnetic field and discuss its implications for the chiral phase transition at finite temperature. The AMM of the quark/antiquark in the pair points in the same direction, as the pair is formed by particles with opposite spins and opposite charges. Hence, the pair has a nonzero magnetic moment that becomes aligned by the external magnetic field and then producing the magnetization of the ground state. This magnetization is reflected in the existence of a second independent condensate. The two condensates contribute to the effective dynamical mass, which is mainly determined by the quark/antiquark pairing in the lowest Landau lever (LLL), resulting in a significant increase in the critical temperature for the chiral restoration, as compared to the case where only the magnetically catalyzed chiral condensate is considered.

\section{Model and Condensates}\label{Sec-2}

Let us consider the following  NJL model of massless quarks in the presence of a constant and uniform magnetic field. The new element of the proposed model is the introduction of a four-fermion channel, with coupling constant $G'$, that becomes relevant only in the presence of a magnetic field,
\begin{eqnarray}\label{lagrangian}
\mathcal{L}&=&\bar{\psi}i\gamma^{\mu}D_{\mu}\psi+\frac{G}{2}[(\bar{\psi}\psi)^2+(\bar{\psi}i\gamma^5\psi)^2]\nonumber
 \\
&+&\frac{G'}{2}[(\bar{\psi}\Sigma^3\psi)^2+(\bar{\psi}i\gamma^5\Sigma^3\psi)^2]
\end{eqnarray}
The new interaction channel naturally emerges using the Fierz identities in the one-gluon-exchange channels of QCD when the rotational symmetry is broken by the magnetic field. Here, $\Sigma^3$ is the spin operator in the direction of the applied field.

Solving the system gap equations in the LLL, we find the condensate solutions 
\begin{equation}\label{sigma}
\overline{\sigma}=G\langle\bar{\psi}\psi\rangle=A\exp{-\left[\frac{2\pi^2}{(G+G')N_cqB}\right]}
\end{equation}
and
\begin{equation}\label{AMM}
\overline{\xi}=G' \langle\bar{\psi}i\gamma^1 \gamma^2\psi\rangle
=A'\exp{-\left[\frac{2\pi^2}{(G+G')N_cqB}\right]}
\end{equation}
with $A=2G\Lambda/(G+G')$ and $A'=2G'\Lambda/(G+G')$. The condensate $\overline{\sigma}$ is associated with the dynamically generated mass and $\overline{\xi}$ with the AMM.

It is worth to underline that the induced condensates (\ref{sigma})-(\ref{AMM}) depend nonperturbatively on the coupling constants and the magnetic field. This behavior reflects the important fact that in a massless theory, chiral symmetry can be only broken dynamically, that is, nonperturbatively. In this result, the LLL plays a special role due to the absence of a gap between it and the Dirac sea. The rest of the LLs are separated from the Dirac sea by energy gaps that are multiples of $\sqrt{2qB}$, and hence do not significantly participate in the pairing mechanism at the subcritical couplings where the magnetic catalysis phenomenon is relevant. Since the dynamical generation of the AMM is produced mainly by the LLL pairing dynamics, one should not expect to obtain a linear-in-B AMM term, even at weak fields, in sharp contrast with the AMM appearing in theories of massive fermions. In the later case, not only the AMM is obtained perturbatively through radiative corrections, but considering the weak-field approximation means first summing in all the LL's, which contribute on the same footing, and then taking the leading term in an expansion in powers of B \cite{Schwinger, Ioffe}.  Notice that such a linear dependence does not hold, even in the massive case, if the field is strong enough to put all the fermions in the LLL \cite{Jancovici}.

\section{Critical Temperature}\label{Sec-3}

The effect of the new condensate $\langle\bar{\psi}i\gamma^1 \gamma^2\psi\rangle$ is to increase the effective dynamical mass of the quasiparticles in the LLL,
 \begin{equation}\label{neweffectivemass}
M_{\xi}=\overline{\sigma}+\overline{\xi}=2\Lambda\exp{-\left[\frac{2\pi^2}{(G+G')N_cqB}\right]}
\end{equation}

In QCD, for fields, $qB \sim \Lambda^2$, the dimensional reduction of the LLL fermions would constraint the LLL quarks to couple with the gluons only through the longitudinal components. Thus, to consistently work in this regime within the NJL model, we should consider that $G' = G$ (see Ref. \cite{NJL-B}  for details).

Because the effective coupling enters in the exponential, the modification of the dynamical mass by the magnetic moment condensate can be significant. Thus, the quasiparticles become much heavier in our model than in previous studies that ignored the magnetic moment interaction. Taking into account that for $qB/\Lambda^2\sim 1$, $\eta \simeq 1$ in $G'=\eta G$, and using the values $G\Lambda^2=1.835$, $\Lambda=602.3$ MeV \cite{G-Interaction}, $N_c=3$ and $q=|e|/3\simeq 0.1$, it was found in \cite{NJL-B} that due to the condensate $\overline{\xi}$ the dynamical mass of the quasiparticles increases sixfold.

Starting from the thermodynamic potential in the condensate phase
\begin{eqnarray}\label{Thermo-Potential-2}
\Omega_0^T(\sigma,\xi)&=&N_cqB\int_{0}^\Lambda \frac{dp_3}{2\pi^2}\left[ \varepsilon_0 +\frac{2}{\beta}\ln \left(1+e^{-\beta\varepsilon_0}\right)\right]\nonumber
 \\
&+&\frac{\sigma^2}{2G}+\frac{\xi^2}{2G'}, \quad \varepsilon_0^2=p_3^2+\sigma^2+\xi^2,
\end{eqnarray}
we can analytically find the critical temperature $T_{C_\chi}$ from 
\begin{eqnarray}\label{Critical-Temperature-1}
\frac{\partial^2\Omega_0^{T_{C_\chi}}}{\partial \overline{\sigma}^2}|_{\overline{\sigma}=\overline{\xi}=0}=-\left [C\int_{0}^\Lambda \frac{dp_3}{p_3}\tanh \left(\frac{\beta_{C_\chi} p_3}{2} \right ) +\frac{1}{G}\right]\nonumber
\end{eqnarray}
\begin{equation}\label{Critical-Temperature}
+\frac{\sigma^2}{2G}+\frac{\xi^2}{2G'}=0
\end{equation}
where $C=N_cqB(G+G')/(2\pi^2G)$.

Doing in (\ref{Critical-Temperature}) the change $p_3 \rightarrow p_3/T_{C_\chi}$, we have
\begin{equation}\label{Critical-Temperature-2}
\int_{0}^{\Lambda/T_{C_\chi}} \frac{dp_3}{p_3}\tanh \left (\frac{p_3}{2} \right ) =\frac{2\pi^2}{(G+G')N_cqB},
\end{equation}
Solving this equation for $T_{C_\chi}$, we get
\begin{equation}\label{Critical-Temperature-3}
T_{C_\chi}=1.16\Lambda \exp -\left [\frac{2\pi^2}{(G+G')N_cqB}\right ]=0.58M_\xi
\end{equation}
The fact that the critical temperature is proportional to the dynamical mass at zero temperature, is consistent with what has been found in other models \cite{Temperature}. In the present case, since the dynamical mass is increased by the AMM, the critical temperature is proportionally increased.
Notice that, we would have arrived at the same result by taking instead the derivative with respect to $\overline{\xi}$. This is a consequence of the proportionality between $\overline{\sigma}$ and $\overline{\xi}$, (i.e. $\overline{\xi}=(G'/G)\overline{\sigma}$), which follows from Eqs. (\ref{sigma}) and (\ref{AMM}). This implies that  the two condensates evaporate at the same critical temperature.
The fact that there exists a unique critical temperature for the evaporation of the two condensates indicates that the condensate $\overline{\xi}$ does not break any new symmetry that was not already broken by the condensate $\overline\sigma$ and the magnetic field, as pointed out above. The simultaneous evaporation of the chiral and magnetic moment condensates has been also reported in the context of lattice QCD \cite{Costa}. 

\section{Conclusion and Discussion}\label{Sec-3}

In the presence of a magnetic field there is no magnetically catalyzed chiral condensate $\langle \overline{\psi}\psi\rangle$ without the simultaneous generation of a second dynamical condensate of the form $\langle \overline{\psi}\Sigma^3\psi\rangle$. The genesis of this phenomenon lies in the fact that the chiral pairs possess net magnetic moments that tend to align with the external magnetic field. The collective effect of these magnetic moments leads to the ground state magnetization and manifests itself as a spin-one condensate
$\langle \overline{\psi}\Sigma^3\psi\rangle$ which enters in the quasiparticle spectrum as an AMM. 

An important effect of $\langle \overline{\psi}\Sigma^3\psi\rangle$ is to increase the effective dynamical mass of the LLL quarks, and consequently the critical temperature of the chiral phase transition. Since the quasiparticles become heavier, compared to the case when the spin-one condensate is ignored, and since they are charged, the electrical conductivity in this case should be much smaller at strong fields. 

The characteristic increase of the critical temperature with an applied magnetic field in the MC$\chi$SB phenomenon is in sharp contrast with the dropping of the temperature with the field found in lattice QCD \cite{QCD-latticeB}. A reconciliation between these apparently contradictory results can be worked out, once the running of the coupling with the magnetic field is incorporated into the analysis \cite{running-coup}. As discussed in  \cite{running-coup}, a strong magnetic field gives rise to an anisotropy of the strong coupling constant. The main effect for the critical temperature is then associated with the running of the parallel coupling $\alpha_s^\|$, which characterizes the interactions in the direction parallel to the field and can be connected to the conventional NJL couplings through the relation $G+G'=4\pi\alpha_s^\|/qB$. The behavior of this coupling with the field is such that the critical temperature ends up decreasing with the field, in agreement with the lattice results. The physical mechanism behind this effect can be traced back to the antiscreenning produced by the quarks confined to the LLL in a strong magnetic field \cite{running-coup}.




\nocite{*}
\bibliographystyle{elsarticle-num}
\bibliography{martin}

\begin{thebibliography}{00}
\bibitem{QCD-latticeB} G. Bali, \textit{et. al}, JHEP 1202, 044 (2012); G. S. Bali, \textit{et. al}, Phys. Rev. D 86, 071502 (2012).


\bibitem{NJL-B} E. J. Ferrer, V. de la Incera, I. Portillo and M. Quiroz, Phys. Rev. D  89, 085034  (2014).

\bibitem{MC} K. G. Klimenko, Z. Phys. C 54, 323 (1992); V. P. Gusynin, V. A. Miransky and I. A. Shovkovy, Phys. Rev. Lett. 73, 3499 (1994).

\bibitem{severalaspects}
C. N. Leung, Y. J. Ng and A. W. Ackley, Phys. Rev. D 54, 4181
(1996); E. J. Ferrer and V. de la Incera, Phys. Rev. D 58, 065008 (1998); V. P. Gusynin, V. A. Miransky and I. A. Shovkovy, Nucl. Phys. B 563, 361 (1999);
E. J. Ferrer and V. de la Incera, Phys. Lett. B 481, 287 (2000); Yu. I. Shilnov, and V. V. Chitov, Phys. Atom. Nucl. 64 (2001) 2051 [Yad. Fiz. 64, 2138 (2001)]; E. J. Ferrer, V. P. Gusynin, and V. de la Incera, Eur. Phys. J. B 33, 397 (2003); N. Sadooghi, A. Sodeiri Jalili, Phys. Rev. D 76, 065013 (2007); E. Rojas, A. Ayala, A. Bashir, and A. Raya, Phys. Rev. D 77, 093004. (2008).

\bibitem{ferrerincera}
E. J. Ferrer and V. de la Incera, Phys. Rev. Lett. 102, 050402 (2009); E. J. Ferrer and V. de la Incera, Nucl. Phys. B 824, 217 (2010).

\bibitem{Schwinger}J. Schwinger, Phys. Rev. 73, 416 (1947).

\bibitem{Ioffe}B. L. Ioffe and A. V. Smilga, Nucl. Phys. 232, 109 (1984).

\bibitem{Jancovici}B. Jancovici Phys. Rev. A 187, 2275 (1969); E. J. Ferrer, V de la Incera, D. Manreza Paret, A. Pérez Martínez; "Anomalous-Magnetic-Moment Effects in a Strongly Magnetized and Dense Medium" in the electronic proceedings of the conference: Compact Stars in the QCD Phase Diagram III (CSQCD III), Guarujá, SP, Brazil (arXiv:1307.5947 [nucl-th]).

\bibitem{G-Interaction}P. Rehberg, S. P. Klevansky and J. H\"{u}fner, Phys. Rev. C 53, 410 (1996).

\bibitem{Temperature} V. P. Gusynin and I. A. Shovkovy, Phys. Rev. D 56, 5251 (1997).

\bibitem{Costa} G. S. Bali, \textit{et. al}, Phys. Rev. D 86, 094512 (2012).

\bibitem{running-coup}
E. J. Ferrer, V. de la Incera and X. J. Wen, \textit{Quark Antiscreening at Strong Magnetic Filed and Inverse Magnetic Catalysis}, arXiv:1407.3503 [nucl-th].






 \end{thebibliography}

\end{document}